\documentclass[%
 reprint,
 amsmath,amssymb,
 aps,
 prb,
]{revtex4-2}

\usepackage{graphicx}%
\usepackage{dcolumn}%
\usepackage{bm}%
\usepackage{hyperref}%

\RequirePackage{hyperref}
\usepackage[utf8]{inputenc} %

\DeclareUnicodeCharacter{03B4}{{$\delta$}}
\graphicspath{ {figures/} }

\begin{document}

\title{The frequency-resolved frozen phonon multislice method and its application to vibrational EELS using parallel illumination}

\author{Paul M. Zeiger}
\email{paul.zeiger@physics.uu.se}
\affiliation{%
  Department of Physics and Astronomy, Uppsala University, P.O. Box 516, 75120 Uppsala, Sweden
}%
\author{J\'{a}n Rusz}%
\affiliation{%
  Department of Physics and Astronomy, Uppsala University, P.O. Box 516, 75120 Uppsala, Sweden
}%

\date{\today}%

\begin{abstract} 
We explore the capabilities of the frequency-resolved frozen phonon multislice method introduced in Phys. Rev. Lett. \textbf{124}, 025501 (2020) to model inelastic vibrational scattering in transmission electron microscopy. We review the method in detail and discuss advantages of using a so called hotspot thermostat instead of the $\delta$-thermostat used in our first report. We apply the method to simulate vibrational electron energy loss spectra of hexagonal boron nitride under plane wave illumination. Simulated spectroscopic information well represents the theoretical phonon bandstructure of the studied material, both in terms of energies as well as polarization vectors of individual phonon modes.
\end{abstract}

\maketitle

\section{Introduction}

A new generation of electron beam monochromators have reduced the energy resolution of electron energy loss spectroscopy (EELS) to a few meV \cite{krivanek_vibrational_2014}, with the record currently being around 4.2~meV \cite{krivanek_progress_2019} and routine measurements delivering around 10-20 meV. This energy scale coincides with the characteristic energies of vibrational excitations in matter, known as phonons. Prior to the advent of  high resolution EELS, the development of aberration correctors pushed the (scanning) transmission electron microscope [(S)TEM] to now routinely allow for sub-{\AA} spatial resolution \cite{haider_electron_1998,batson_sub-angstrom_2002,akashi_aberration_2015,sawada_resolving_2015,morishita_resolution_2016,morishita_resolution_2018}.

Since its inception, high resolution EELS instrumentation enabled a number of exciting experiments, especially, on graphene-type 2d-materials, multilayers and organic molecules. These specimen tend to have rather large vibrational energies due to the small mass of their constituent atoms, which facilitates the acquisition of vibrational EELS spectra. On these systems the capabilities of STEMs of combining excellent energy and spatial resolution was demonstrated. For example the phonon dispersion of hBN and graphene were measured from areas of nanometer scale \cite{hage_nanoscale_2018,senga_position_2019}. In these experiments $\mathbf{k}$-space resolution is retained by beams with a small convergence angle. For the highest spatial resolution, the convergence angle needs to be opened up to a few tens of milliradians, in order to achieve atomic resolution. In this type of experiments it was demonstrated, that both off- and on-axis EELS exhibit atomic scale changes of the vibrational EELS spectrum \cite{hage_phonon_2019, venkatraman_vibrational_2019}. The technique was pushed to the extreme by measuring changes in the vibrational scattering induced by single silicon impurity in a monolayer of graphene \cite{hage_single-atom_2020}.

It was furthermore shown that the EELS signal is sensitive to two different regimes of vibrational scattering, so-called dipolar and impact scattering \cite{rez_damage-free_2016,dwyer_electron-beam_2016}. The former appears only in polar materials such as hexagonal boron-nitride (hBN) and is associated with an dipolar interaction between the beam electron and the time-dependent polarization of the sample, whereas the latter appears in all specimen and can be attributed to scattering off of the ionic cores of atoms. Both scattering regimes have been combined to map out the surface and bulk modes of a magnesium oxide nanocube \cite{lagos_mapping_2017}.

Instrumental development has not stopped with aberration correctors and better monochromators: using a new generation of detectors with much reduced noise levels, angle resolved vibrational EELS spectra were measured on multilayer hBN in parallel \cite{plotkin-swing_hybrid_2020}. These spectra allow to effectively capture the phonon dispersion of the material along a certain direction in reciprocal space.

From a theoretical point of view vibrational EELS spectra are often treated using a first Born approximation of the scattering process \cite{senga_position_2019, nicholls_theory_2019,hage_single-atom_2020}. The basic theory for this approach was laid out by Van Hove more than half a century ago \cite{van_hove_correlations_1954}. The popularity of this approach is its computational simplicity: the inelastic scattering cross section is expressed as a sum of transition matrix elements over phonon modes. For small systems, this sum can be evaluated using accurate phonon modes obtained from density functional theory (DFT). One thereby completely ignores, however, the deformation of the electron beam due to the strong elastic interaction with the specimen and it is therefore strictly speaking not a well justified model for specimen of finite thickness. Within Van Hove theory, it is  for qualitative purposes often sufficient to compare the EELS just to the projected phonon density of states (PDOS) as shown in the supplementary material in Ref.~\cite{hage_single-atom_2020}.

Another approach was taken by Dwyer, who used a single inelastic scattering theory to investigate possible obtainable spatial resolution for different experimental conditions on a multilayer hBN sample \cite{dwyer_prospects_2017}. In this approach, an elastic wavefunction is propagated through the specimen using the multislice method and at each atomic layer the inelastic wave associated with a certain energy is accumulated by evaluating inelastic transition matrix elements corresponding to those transitions. This approach describes elastic and inelastic scattering processes, and includes both dipolar as well as impact scattering. A disadvantage is, that the approach becomes quickly computationally expensive as the unit cell and number of considered modes grow, because it requires explicit knowledge of all vibrational modes.

The frozen phonon multislice (FPMS) method introduced by Loane et al. is frequently used to include the effect of thermal diffuse scattering (inelastic phonon scattering) into diffraction patterns and STEM images \cite{loane_thermal_1991}. It is flexible and highly accurate \cite{lebeau_quantitative_2008}. In the FPMS method the thermally averaged diffraction pattern $I_{\mathrm{FPMS}} (\mathbf{q_{\perp}}, \mathbf{r}_{\mathrm{b}})$ is expressed as an incoherent average of $N$ reciprocal space crystal exit plane wave functions $\Psi\left(\mathbf{q}_{\perp}, \mathbf{r}_{\mathrm{b}}, \mathbf{R}_{n} \right)$ computed from atomic configurations $\left\lbrace \mathbf{R}_n \right\rbrace_{n=1}^{N}$ of the vibrating structure model (henceforth called "snapshot"), i.e.,
\begin{align}
    I_{\mathrm{FPMS}} (\mathbf{q}_{\perp}, \mathbf{r}_{\mathrm{b}}) = & {} \frac{1}{N} \sum_{n=1}^{N} \left|\Psi\left(\mathbf{q}_{\perp}, \mathbf{r}_{\mathrm{b}}, \mathbf{R}_{n} \right) \right|^2, \label{eq:FPMS_incohavg}
\end{align}
where $\mathbf{q_{\perp}}$ is the momentum transfer in the diffraction plane and $\mathbf{r}_{\mathrm{b}}$ is the STEM beam position. The exit wave function $\Psi\left(\mathbf{q}_{\perp}, \mathbf{r}_{\mathrm{b}}, \mathbf{R}_{n} \right)$ is computed using the elastic multislice method \cite{cowley_scattering_1957}. The FPMS method allows for great freedom in how the snapshots $\mathbf{R}_{n}$ are generated. The simplest way is to assume no correlation in the atomic motion, i.e., an Einstein model, and randomly displace atoms from their equilibrium position, as it is done in the majority of calculations. More sophisticated models take correlations in atomic motion into account by means of molecular dynamics simulations \cite{mobus_influence_1998,a._muller_simulation_2001,biskupek_evaluation_2007,rother_statistics_2009,aveyard_modeling_2014,lofgren_influence_2016,pohl_atom_2017,krause_using_2018}. The frozen phonon model was conceived from the semi-classical picture, that the electron passes the the TEM specimen during a time interval, which is much shorter than the characteristic time scale of atomic vibrations, and that the time, which passes between two consecutive beam electrons enter the sample, is much longer than the characteristic time scale of atomic vibrations \cite{wang_elastic_1995}. It was further theoretically justified on the basis of quantum mechanical considerations \cite{wang_elastic_1995,wang_`frozen-lattice_1998,van_dyck_is_2009,niermann_scattering_2019}. A numerically equivalent expression to equation \ref{eq:FPMS_incohavg} can be derived from a Born-Oppenheimer approximation of the Many-Body Schrödinger equation, an approach known as the quantum excitations of phonons (QEP) model \cite{forbes_quantum_2010}.

The FPMS methods has, however, a major shortcoming: it cannot provide spectral information, since the thermally averaged intensity contains contributions of all vibrational frequencies. We recently extended the FPMS method by adding the missing frequency dependency \cite{zeiger_efficient_2020}. Our method leverages in its current form non-equilibrium Molecular Dynamics (MD) simulations, in which specific thermostats enforce an excitation of a certain range of vibrational frequencies. We shall call this new method the ``frequency resolved frozen phonon multislice'' (FRFPMS) method from here on.

In this article, we will provide a more detailed insight into the machinery and properties of the FRFPMS simulations using hBN as the example system. We show how the choice of the thermostat in the underlying MD simulation affects the vibrational power spectrum of frequencies in the resulting atom trajectories and consider the vibrational signal as a function of momentum transfer in the diffraction plane for planar wave illumination. We compare the results to calculations of the phonon band structure and of the PDOS of the hBN model. Additionally we compare integrated off-axis spectra to the PDOS.%

\section{Methods}

This section is organized in the following way: first we describe the FRFPMS method in detail. Then we shift our focus to the key ingredients of the method, i.e., the elastic multislice method for obtaining crystal exit wave functions of fast electrons and MD simulations for the generation of atomic configurations. Lastly, we outline the procedure to obtain vibrational power spectra from MD trajectories and describe our phonon calculations. We pay close attention to name and explain all the parameters and settings of our simulations in this section.

\subsection{FRFPMS method}

We have outlined the general procedure of the FRFPMS method earlier \cite{zeiger_efficient_2020}, but describe it here for the sake of completeness and detail. The core idea of the FRFPMS method is to select a set of frequencies $\omega_i$, $i=1,\ldots, N_{\mathrm{bin}}$, within the range of the vibrational frequencies of the material under study, henceforth just called frequency or energy "bins", and to perform essentially one full FPMS simulation for each of these frequency bins using a set of $N$ atomic configurations $\left\lbrace \mathbf{R}_n[\omega_i] \right\rbrace_{n=1}^{N}$, the snapshots, corresponding to the selected frequency $\omega_i$. We sample the atomic configurations from a MD simulation, in which predominantly the selected frequency $\omega_i$ is excited.

In this way, we obtain $N_{\mathrm{bin}}\times N$ crystal exit plane wave functions $\Psi\left(\mathbf{q}_{\perp}, \mathbf{r}_{\mathrm{b}}, \mathbf{R}_n[\omega_i]\right)$. For each bin, we compute the incoherent and coherent intensities, $I_{\mathrm{incoh}}$ and $I_{\mathrm{coh}}$, respectively, according to Ref.~\cite{forbes_quantum_2010}, i.e.,
\begin{equation}
\begin{aligned}
    I_{\mathrm{incoh}}(\mathbf{q}_{\perp}, \mathbf{r}_{\mathrm{b}}, \omega_i) = & {} \frac{1}{N} \sum_{n=1}^{N} \left|\Psi\left(\mathbf{q}_{\perp}, \mathbf{r}_{\mathrm{b}}, \mathbf{R}_n[\omega_i] \right) \right|^2 \\
    = & {} \left\langle \left| \Psi\left(\mathbf{q}_{\perp}, \mathbf{r}_{\mathrm{b}}, \mathbf{R}[\omega_i]\right) \right|^2 \right\rangle_{N} \\
    I_{\mathrm{coh}}(\mathbf{q}_{\perp}, \mathbf{r}_{\mathrm{b}}, \omega_i) = & {} \left|\frac{1}{N}  \sum_{n=1}^{N} \Psi\left(\mathbf{q}_{\perp}, \mathbf{r}_{\mathrm{b}}, \mathbf{R}_n[\omega_i]\right) \right|^2 \\
    = & {} \left| \left\langle \Psi \left(\mathbf{q}_{\perp}, \mathbf{r}_{\mathrm{b}}, \mathbf{R}[\omega_i]\right) \right\rangle_{N} \right|^2,
\end{aligned}
\label{eq:Icoh_Iincoh}
\end{equation}
where $\langle\;\cdot\;\rangle_{N}$ denotes the arithmetic mean over uncorrelated snapshots $\left\lbrace \mathbf{R}_n[\omega_i] \right\rbrace_{n=1}^{N}$. The vibrational intensity $I_{\mathrm{vib}}(\mathbf{q}_{\perp}, \mathbf{r}_{\mathrm{b}}, \omega_i)$ is given by the difference between these intensities, i.e.,
\begin{equation}
\begin{aligned}
      & I_{\mathrm{vib}}(\mathbf{q}_{\perp}, \mathbf{r}_{\mathrm{b}}, \omega_i) \\
     = {} & I_{\mathrm{incoh}}(\mathbf{q}_{\perp}, \mathbf{r}_{\mathrm{b}}, \omega_i) - I_{\mathrm{coh}}(\mathbf{q}_{\perp}, \mathbf{r}_{\mathrm{b}}, \omega_i) \\
     = {} & \left\langle \left| \Psi\left(\mathbf{q}_{\perp}, \mathbf{r}_{\mathrm{b}}, \mathbf{R}[\omega_i]\right) \right|^2 \right\rangle_{N} \\
     & - \left| \left\langle \Psi \left(\mathbf{q}_{\perp}, \mathbf{r}_{\mathrm{b}}, \mathbf{R}[\omega_i]\right) \right\rangle_{N} \right|^2.
\end{aligned}
\label{eq:Ivib}
\end{equation}
It is worth pointing out that in a statistical sense, the vibrational intensity $I_{\mathrm{vib}}(\mathbf{q}_{\perp}, \mathbf{r}_{\mathrm{b}}, \omega_i)$ is the variance of the reciprocal space crystal exit plane beam wave functions. Furthermore, it becomes clear from this form of the vibrational intensity, that it is a function of the two dimensional momentum transfer $\mathbf{q}_{\perp}$ in the diffraction plane, the two dimensional position vector of the STEM probe $\mathbf{r}_{\mathrm{b}}$ and the frequency $\omega_i$. Thus the FRFPMS method yields a five dimensional data hyper cube $I_{\mathrm{vib}}(\mathbf{q}_{\perp}, \mathbf{r}_{\mathrm{b}}, \omega_i)$. STEM-EELS experiments can similarly give access to spatial, momentum and energy dimensions. In order to make FRFPMS data comparable to experiment we need to take two further steps.

The first step signifies the FRFPMS's interpretation of the inelastic scattering process: inelastic scattering at energy loss $\Delta E$ is assumed to be equivalent to the incoherent part of repeated elastic scattering on atomic configurations of the specimen, which vibrates predominantly with frequency $\omega = \Delta E / \hbar$, i.e., we take
\begin{equation}
\begin{aligned}
    I_{\mathrm{vib}}(\mathbf{q}_{\perp}, \mathbf{r}_{\mathrm{b}}, \omega_i) = {} & I_{\mathrm{vib}}(\mathbf{q}_{\perp}, \mathbf{r}_{\mathrm{b}}, \Delta E_i/\hbar) \\
    \equiv {} & I_{\mathrm{vib}}(\mathbf{q}_{\perp}, \mathbf{r}_{\mathrm{b}}, \Delta E_i).
\end{aligned}
\end{equation}

Secondly, the spectral information is in experiments often, but not always, integrated over an EELS spectrometer entrance aperture $\Omega(\mathbf{q}_{\perp})$ centered around some adjustable momentum transfer $\mathbf{q}_{\perp}$ in the diffraction plane. In order to compare to such an experiment, we need to compute the integral
\begin{equation}
    I_{\mathrm{vib}}(\mathbf{r}_{\mathrm{b}}, \Omega({\mathbf{q}_{\perp}}), \Delta E) = \int\limits_{\Omega(\mathbf{q}_{\perp})} \mathrm{d}\mathbf{q}_{\perp}' \; I_{\mathrm{vib}}(\mathbf{q}_{\perp}', \mathbf{r}_{\mathrm{b}}, \Delta E). \label{eq:Ivib_spectra}
\end{equation}
We will call $I_{\mathrm{vib}}(\mathbf{r}_{\mathrm{b}}, \Omega({\mathbf{q}_{\perp}}), \Delta E)$ for a particular choice of $\mathbf{r}_{\mathrm{b}}$ and $\Omega({\mathbf{q}_{\perp}})$ often just the "vibrational EELS signal" or "EELS" as in "electron energy loss spectrum".

We should mention at this point, that it is still under investigation, how the FRFPMS's interpretation of the scattering process connects with other theories of vibrational EELS. We can, however, state already at this point some limitations of the model. For example, the model will incorrectly bin the majority of multiple inelastic scattering processes, since there is no interaction term between energy bins, which would allow say a doubly scattered electron to be accounted for in an energy bin corresponding to twice the energy loss. Conversely, as long as single inelastic scattering dominates, i.e., for specimen thicknesses smaller than the mean free path length of phonon scattering, typically on the order of tens to several hundreds of nanometer \cite{hall_effect_1965}, we expect the model to be quite accurate, supported by the results we obtain and their favourable comparison with experiment. Another inaccuracy of the FRFPMS method stems from the necessarily finite width of frequency excitation in the generation of snapshots. This implies, that within each energy bin, multiple phonon modes are excited and therefore considered coherently, in contrast to them being actually incoherent due to differing associated energy losses. This is, however, inherent to the FPMS method in general and FRFPMS actually reduces the issue by splitting the pool of all phonon modes into narrower energy bins.

The major advantage of the FRFPMS procedure over other approaches for vibrational EELS simulations is, that we do not require explicit knowledge of the phonon modes of the considered specimen: MD simulations have a computational complexity of order $N_{\mathrm{at}}$, where $N_{\mathrm{at}}$ is the number of atoms in the simulation, as long as empirical, finite range potentials are considered, whereas the eigenvalue problem involved in obtaining the phonon modes has a computational complexity of order $N_{\mathrm{at}}^3$. In the next paragraph, we will describe the structure model, which we have used for our simulations.

\subsection{The structure model}

We use a simulation box of 24$\times$10$\times$46 hexagonal unit cells of AA'-stacked hBN in the $x$, $y$, and $z$-directions containing in total $N_{\mathrm{at}}=22080$ atoms. The multislice calculations require an orthogonal simulation box and we have thus chosen an orthogonal super cell already at the MD level of the calculations. The crystallographic $c$-axis is aligned with the $z$-direction of the box. The relaxed orthogonal simulation box has a size of 52.01~{\AA} $\times$ 25.02~{\AA} $\times$ 149.71~{\AA}. We note that the shape of the box, i.e., having a side ration of roughly 2:1 in $x/y$-dimensions, is sub-optimal for multislice calculations, where one usually opts for lateral dimensions closer to a 1:1 ratio. The choice made here is motivated by a follow-up work, which considers vibrational STEM-EELS simulations for hBN with a defect plane \cite{hBN_APB_2021}. There, due to periodic boundary conditions, the super cell needs to accommodate two defect planes and, in an effort to save computational resources, a 2:1 ratio of $x/y$-dimensions becomes the natural choice. A defect-free super cell of the same starting dimensions has been constructed as a reference system, and that is the structure model used here. Here we can highlight an additional advantage of the FRFPMS method, i.e., that we can use the same MD trajectories in both works and need to only consider a different incident beam in the multislice simulations, which we will describe in the following paragraph.

\subsection{Elastic multislice method}

In order to compute the reciprocal space crystal exit plane wave function $\Psi\left(\mathbf{q}_{\perp}, \mathbf{r}_{\mathrm{b}}, \mathbf{R}\right)$ of a given assembly of atoms $\mathbf{R}$, one needs to solve the Schrödinger equation for a high-energy electron along a certain direction. This task can be efficiently solved by means of the multislice method \cite{cowley_scattering_1957}, in which the scattering potential is divided into many slices perpendicular to the beam direction and such that it is weak within each slice. An initial electron beam wave function is then transformed into the crystal exit plane wave function by successive transmission and propagation steps for each slice of the potential. In practice, one prepares the initial beam wave function according to the desired beam specifications. We use here the elastic multislice method as implemented in the software package \texttt{DrProbe} within the projected potentials approximation \cite{barthel_dr_2018}.

We set the beam acceleration voltage to 60~keV, which is a common value in vibrational EELS experiments. We choose furthermore a parallel illumination, i.e., the convergence semi-angle is 0~mrad and the initial wave function is given by a plane wave. Such a wave fills the entire simulation box uniformly and the wave function is therefore unchanged by real space shifts. Thus the data set, which we consider in this work, does not depend on the beam position $\mathbf{r}_{\mathrm{b}}$, i.e., we have in the following $I_{\mathrm{vib}}(\mathbf{q}_{\perp}, \mathbf{r}_{\mathrm{b}}, \omega_i) \equiv I_{\mathrm{vib}}(\mathbf{q}_{\perp}, \omega_i)$. As mentioned earlier, we need to perform $N_{\mathrm{bin}}\times N$ multislice calculations, i.e., one for each energy bin and snapshot within the bin. The lateral dimensions of the computational grid for the multislice calculations is chosen to be $1008\times480$. The reciprocal space sampling density is 0.12~\AA$^{-1}$ and 0.25~\AA$^{-1}$ in $q_x$ and $q_y$-directions, respectively. In total 750 slices of a thickness 0.2~\AA{} have been generated. We will now turn towards describing the MD calculations, which produce the frequency-dependent snapshots.

\subsection{Molecular Dynamics}
\label{sec:methods_MD}

MD simulations are a standard tool in modern computational materials science. The core of the technique is to integrate equations of motion for a collection of atoms in time in order to sample a certain thermodynamical ensemble. For example the time integration of Newton's equations of motion results in dynamics, which correspond to a constant number of atoms $N$, constant volume $V$ and constant energy $E$ thermodynamic ensemble, also known as $NVE$ or microcanonical ensemble. The experimental reality, however, is usually not well described by the microcanoncial ensemble, since most systems interact in some way with other systems. Thermostats can be used in Molecular Dynamics simulations to enforce other thermodynamical conditions. For example, one can apply a so-called Langevin thermostat to a $NVE$ simulation, which changes the equations of motion to the Langevin equation \cite{schneider_molecular-dynamics_1978}, i.e.,
\begin{equation}
    m_n \, \mathbf{a}_n(t) = \mathbf{F}_n(\mathbf{r}(t)) - \frac{m_n}{\tau} \mathbf{v}_n(t) + \mathbf{G}_{n}(t), \label{eq:Langevin_equation}
\end{equation}
where $m_n$ is the mass, $\mathbf{v}_n$ the velocity, and $\mathbf{a}_n$ the acceleration of the $n$-th atom. $\mathbf{F}_n(\mathbf{r}(t))$ is the force due to the potential acting on the $n$-th atom. The random force $\mathbf{G}_{n}(t)$ is uncorrelated in time, has zero mean over time, and is proportional to $\sqrt{\frac{m_n k_{\mathrm{B}} T}{\tau \Delta t}}$, where $k_{\mathrm{B}}$ is the Boltzmann constant, $\Delta t$ is the time step and $T$ the desired temperature. The damping parameter $\tau$ regulates how aggressively the thermostat acts and how close the resulting dynamics is to pure $NVE$ dynamics, which is the limit of $\tau \rightarrow \infty$. In principle, Langevin thermostatting generates a $NVT$ or canonical ensemble for any finite damping $\tau$, however only after very long time \cite{hunenberger_thermostat_2005}.

The Langevin thermostat, which we have just considered, does not provide for the ability to selectively excite certain frequencies in the MD simulation and we need therefore a different thermostatting scheme. The non-Markovian generalized Langevin equation of motion can be tailored to a wide range of applications \cite{ceriotti_colored-noise_2010}. Among these, frequency dependent heating of normal modes, the so-called $\delta$-thermostat \cite{ref_special:ceriotti_delta-thermostat:_2010}, and selective heating of normal modes atop a white-noise baseline thermostat, the so-called hotspot thermostat \cite{dettori_simulating_2017}. These thermostats heat artificially those vibrational modes, whose frequencies lie within a narrow range of frequencies $\Delta \omega$ around a chosen peak frequency $\omega_0$. Modes, whose frequencies lie outside of $\Delta \omega$, are kept effectively "frozen" in the case of the $\delta$-thermostat or at a base temperature $T_{\mathrm{base}}$ in the case of the hotspot thermostat.

In our first, proof-of-concept report of the FRFPMS method \cite{zeiger_efficient_2020}, we employed the $\delta$-thermostat, whose frequency width is fixed at relative value of $\Delta \omega/\omega_0 = 0.01$, which translates into an average energy resolution of around 1~meV. We will show below, that this narrow frequency width may present certain difficulties for finite-size simulation boxes. Here, we consider as a novelty mainly results from MD simulations using a hotspot thermostat.

We have performed a variety of MD simulations using different thermostats in order to signify the effect of the choice of thermostat on FRFPMS simulations: $NVE$ simulations at an average kinetic energy consistent with a temperature of around 300~K were performed in order to establish a base-line vibrational power spectrum, resulting from the force-field. In another MD simulation, we have applied a Langevin thermostat according to equation~\ref{eq:Langevin_equation} for two different settings of $\tau$, namely 0.5~ps and 0.05~ps. $\delta$-thermostat simulations are performed for a grid of $N_{\mathrm{bin}}=17$ peak frequencies $\omega_0$ from 11.5~THz to 51.5~THz in steps of 2.5~THz. The same grid is used for hotspot simulations with $\Delta \omega = 2.5$~THz. Hotspot simulations with $\Delta \omega = 1.0$~THz are run for a grid of $N_{\mathrm{bin}}=48$ peak frequencies from 5~THz to 52~THz in steps of 1.0~THz. The peak frequency of the $\delta$- and hotspot thermostats is set to 300~K. The parameters $T_{\mathrm{base}}$, $1/\gamma_{\mathrm{base}}$, and $1/\gamma$ of the hotspot thermostat were set to the values 0.0~K, 0.1~ps and 0.5~ps, respectively. The necessary input matrices for the generalized Langevin dynamics can be obtained from a web repository \cite{gle4md_web}.

The interatomic potential for hBN is divided into interlayer and intralayer interactions, which are described by a so-called extended Tersoff potential and an intralayer potential specifically optimized for bulk hBN, respectively \cite{los_extended_2017,ouyang_mechanical_2020}. Partial charges on B and N atoms are accounted for by a shielded Coulomb potential \cite{maaravi_interlayer_2017}. All MD simulations used an integration time step of $0.5$~fs and were run for a total of 0.25~ns, that is 5$\cdot$10$^5$ time steps, with the \texttt{LAMMPS} MD software \cite{plimpton_fast_1995,lammpsweb}. The first 0.025~ns are discarded to allow for a steady state to be reached, before any output is taken. We save snapshots about every 1~ps, i.e., every 2000 timesteps, yielding in total 225 snapshots.

It is those 225 snapshots, over which the averages in the evaluation of incoherent and coherent intensities, i.e., $I_\text{incoh}$ and $I_\text{coh}$ (see Eqn.~(\ref{eq:Icoh_Iincoh})), are performed. We have observed that the vibrational intensity $I_\text{vib}$, which is their difference (Eqn.~(\ref{eq:Ivib})), is well converged everywhere in the diffraction plane for this number of snapshots, except at the Bragg spots, where $I_\text{vib}$ represents a small difference of large numbers. For this reason, we will focus on extracting physical information from regions avoiding the intense Bragg spots in the text below.

\subsection{Computation of vibrational power spectra}

Within the harmonic approximation, the vibrational density of states (VDOS) is the Fourier transform of the velocity autocorrelation function of a $NVE$ MD simulation \cite{lee_ab_1993,carreras_dynaphopy_2017}, i.e.,
\begin{equation}
    g(\omega) = \int_{-\infty}^{\infty} \mathrm{d}t \; \frac{\sum_{n=1}^{N_{\mathrm{at}}} m_n \left\langle \mathbf{v}_n(t) \mathbf{v}_n(0) \right\rangle}{\sum_{n=1}^{N_{\mathrm{at}}} m_n \left\langle \mathbf{v}_n(0) \mathbf{v}_n(0) \right\rangle} \exp(i\omega t), \label{eq:VDOS_from_MD}
\end{equation}
where $\mathbf{v}_n(t)$ is the velocity of the $n$-th atom with mass $m_n$ at time $t$ and $N_{\mathrm{at}}$ is the number of atoms. The velocity autocorrelation function is given by
\begin{equation}
    \left\langle \mathbf{v}_n(t) \mathbf{v}_n(0) \right\rangle = \lim\limits_{\Delta\rightarrow \infty} \frac{1}{\Delta}\int_{0}^{\Delta} \mathrm{d}t' \; \mathbf{v}_n(t+t') \mathbf{v}_n(t').
\end{equation}
In thermal equilibrium, the normalization term in equation~(\ref{eq:VDOS_from_MD}) evaluates to
\begin{equation}
    \sum_{n=1}^{N_{\mathrm{at}}} m_n \left\langle \mathbf{v}_n(0) \mathbf{v}_n(0) \right\rangle = 3N_{\mathrm{at}}k_{\mathrm{B}} T.
\end{equation}
Since the MD simulation is run at some non-zero temperature, the VDOS computed in this way includes temperature-dependent anharmonic effects on the VDOS.

Thermostats modify the equations of motions and thus change the ensemble being sampled by the MD trajectory. Computing $g(\omega)$ according to equation~(\ref{eq:VDOS_from_MD}), will strictly speaking not be the VDOS of the system anymore, but include the action of the thermostat. For this reason, we use in this paper a different terminology and compute what we call the ``vibrational power spectrum'' (VPS) of a MD trajectory, i.e,
\begin{equation}
    \mathrm{VPS}(\omega) = \int_{-\infty}^{\infty} \mathrm{d}t \; \sum_{n=1}^{N_{\mathrm{at}}} \left\langle \mathbf{v}_n(t) \mathbf{v}_n(0) \right\rangle \exp(i\omega t).
\end{equation}
Note how we omit the normalization by the zero-lag autocorrelation. Instead, we choose the normalization on a case-by-case basis. A computationally more efficient yet mathematically equivalent approach is, to compute the square of the Fourier transform of the velocity trajectory, i.e.,
\begin{equation}
\begin{aligned}
    \mathrm{VPS}(\omega)
    = {} & \sum_{n=1}^{N_{\mathrm{at}}} m_n \left| \int_{-\infty}^{\infty}  \mathrm{d}t \; \mathbf{v}_n(t) \exp(i\omega t)\right|^2 \\
    = {} & \sum_{n=1}^{N_{\mathrm{at}}} m_n \left| \mathbf{v}_n(\omega) \right|^2
\end{aligned}
\end{equation}
We use this method as implemented in the \texttt{pwtools} software package \cite{pwtoolsgithub}. We sample all VPSs every 5~ps from the MD trajectories. In the remainder of the text, we will use energy units instead of frequency units. The conversion factor from frequency units (THz) to energy units (meV) is about 4.13~meV/THz.

\subsection{Phonon calculations}

In infinite pure crystals, vibrational excitations are described in terms of phonons and the VDOS is then often called PDOS for these systems. We consider here also phonon properties, such as the phonon dispersion and PDOS computed from the MD force field, in order to compare them with the information contained in the FRFPMS data. Note, that we will differentiate in the following between the VPS, which is computed from a MD simulation, and the PDOS, which is computed directly from the force field.

In order to obatin the phonon dispersion and PDOS, we compute second-order force constants using a finite displacement method and diagonalize subsequently the dynamical matrix obtained from the force constants. To this end, we use the software packages \texttt{phonopy}, \texttt{phonolammps} and \texttt{LAMMPS} in conjunction \cite{phonopygithub,togo_first_2015,phonolammpsgithub,plimpton_fast_1995,lammpsweb}. The super cell consists of $8\times8\times6$ unit cells of hBN in AA'-stacking order for these calculations in order to fully contain the potentials up to their cutoff distances within the cell (all possible neighbor interactions are thus considered). The super cell is first structurally relaxed in order to find the minimum energy and to minimize forces on all atoms using \texttt{LAMMPS}. Then \texttt{phonolammps} is used to obtain the force constants of the system, which are in turn used by \texttt{phonopy} for the calculation of the PDOS and phonon dispersion. The sampling mesh in reciprocal space is $46\times46\times15$ for the calculation of the PDOS and the phonon dispersion was evaluated at 101 points along each of the segments of the path in reciprocal space. For all calculations, the symmetry tolerance parameter of \texttt{phonopy} is fixed at $10^{-8}$.

\section{Results and Discussion}

In line with the goals set out in the introductory section, this section is organized as follows: first we present the phonon band structure and PDOS of hBN computed from the chosen force field. Thereafter we show how the choice of the thermostat influences the VPS of the computed MD trajectories and compare these with the PDOS. We then compare off-axis spectra with the PDOS, consider thickness effects and shed light on the shape of the vibrational signal as a function of the detector position in the diffraction plane. %
Finally, we compare the distribution of the vibrational signal in the diffraction plane with the phonon dispersion along two distinct paths in reciprocal space.

\begin{figure}
    \centering
    \includegraphics[width = \linewidth]{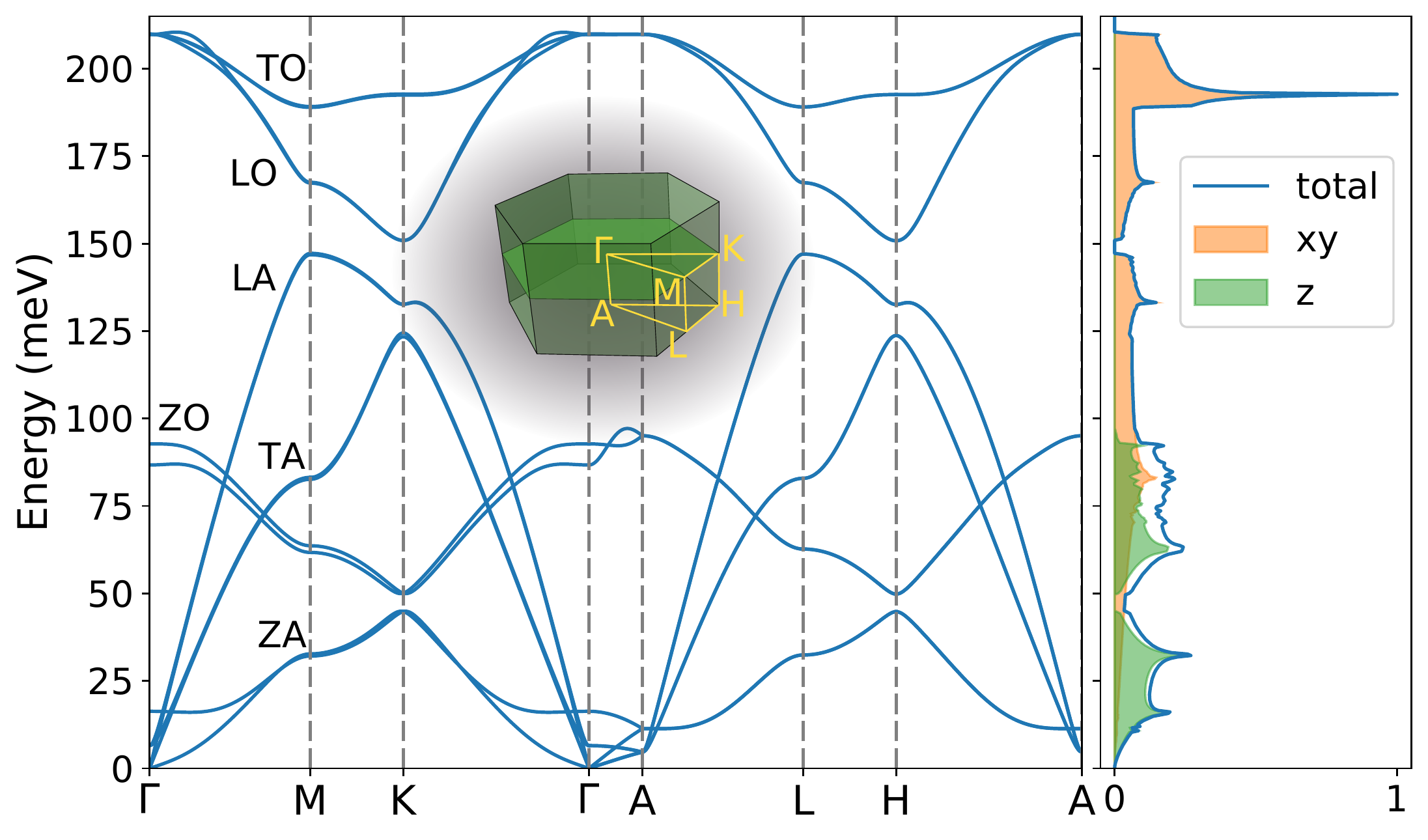}
    \caption{Plot of the phonon dispersion and PDOS of bulk hBN. The inset shows the path in the first Brillouin zone, along which we plot the phonon dispersion. The contribution of the $x/y$- and $z$-projected PDOS to the total PDOS is separately indicated. These results are computed from the second-order force constants of the MD force field using the \texttt{phonolammps} and \texttt{phonopy} software packages.}
    \label{fig:Figure_banddos}
\end{figure}

\subsection{Phonon dispersion of our hBN model}
In Figure~\ref{fig:Figure_banddos} we display a plot of the phonon dispersion of bulk hBN along the indicated path in the first Brillouin zone and the total PDOS. The PDOS is dominated by in plane acoustical (ZA) modes below about 50 meV and exhibits two main peaks in this region. The main contributions between 50 and around 150 meV are transverse acoustical (TA), out-of-plane optical (ZO) and longitudinal acoustical (LA) modes. At higher energies, the main contributions are longitudinal optical (LO) and transversal optical (TO) modes. The TO modes are thereby mainly responsible for the large peak in the PDOS around 190~meV. Our hBN shows furthermore a rather clear separation between the energies of $x/y$- (in-plane) and $z$-vibrations (out-of-plane) as the respective projections of the PDOS indicate. Vibrations in $z$-direction are predominant for energies below around 50~meV and $x/y$-vibrations dominate the PDOS above 100~meV, with the intermediate energies allowing for both $x/y$- and $z$-vibrations.

Naturally the accuracy of the interatomic force field in MD simulations influences the accuracy of spectra computed by the FRFPMS method. Ouyang et al. obtained a similar dispersion albeit using a different intralayer potential for bulk hBN \cite{ouyang_mechanical_2020}. In comparison with experiment, both their and our combination of inter- and intra-layer potentials overestimate the energies of longitudinal optical (LO) and transverse optical (TO) modes, which is also noted by other authors \cite{sevik_characterization_2011}. This overestimation will influence FRFPMS spectra and needs to be taken into account, when comparing them to experimental EELS spectra.

It is worth pointing out, that the PDOS neglects temperature-dependent anharmonic effects. These effects are present in MD calculations and influence in turn the VPS obtained from such calculations. Therefore, anharmonic effects find their way also into the FRFPMS results, but a more detailed study of these effects is outside the scope of this work. We will, however, see that the influence of anharmonic effects is small at a temperature of 300~K for our system.

Neglecting anharmonic effects, vibrational EELS experiments should give us access to the phononic properties shown in Figure~\ref{fig:Figure_banddos}, even at a local level. We will show that the FRFPMS method, which simulates such experiments, indeed carries this information into angle-resolved spectra and, for a carefully chosen experimental geometry, it allows its extraction. There are certain limitations with the current implementation of the FRFPMS method, which we will need to work around or understand before we can attempt such comparisons. Two of such limitations stem from the MD simulations and are to be discussed here in detail.

\begin{figure}
    \centering
    \includegraphics[width = \linewidth]{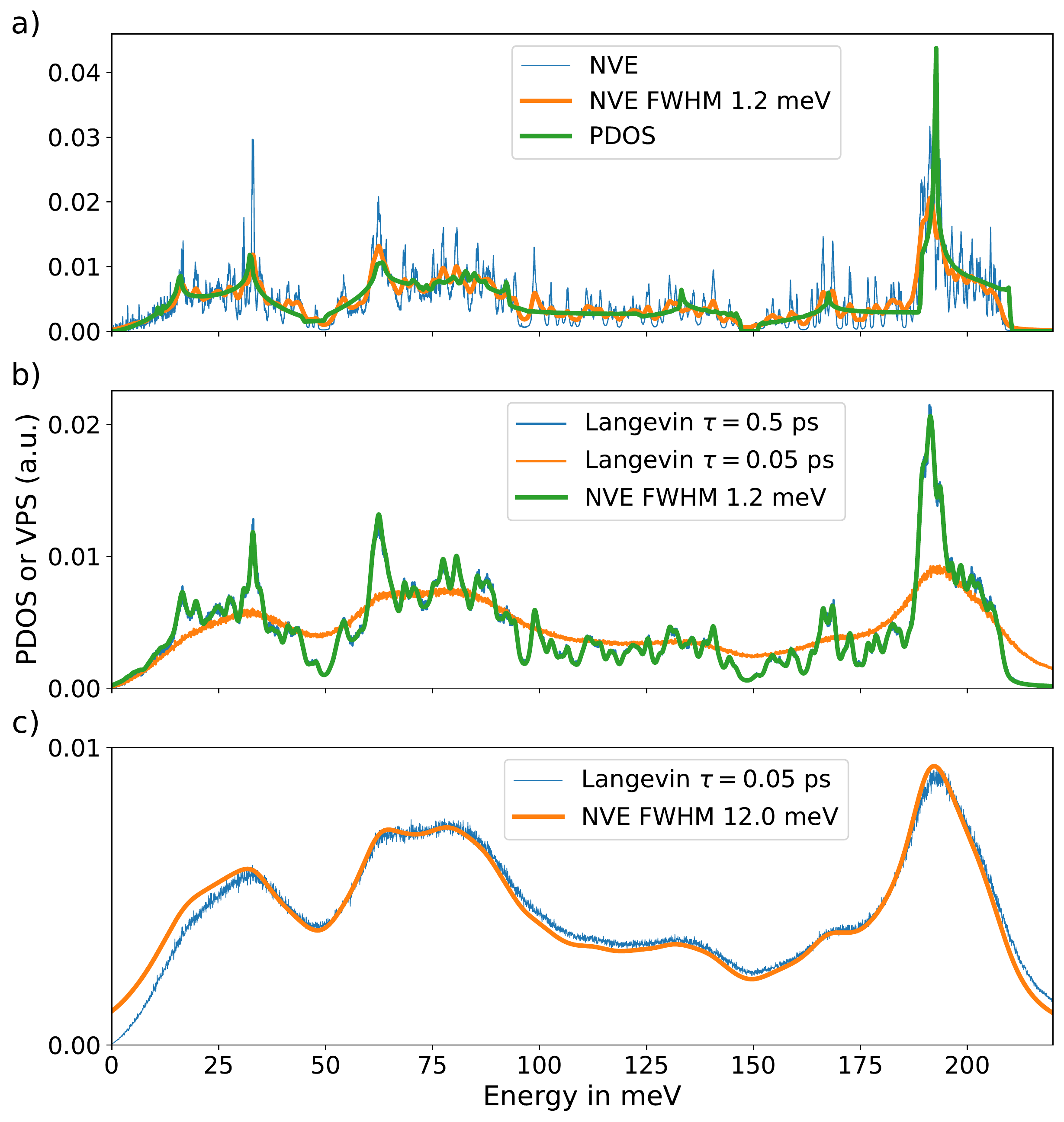}
    \caption{Calculated PDOS and VPS of hBN. Panel a) shows a comparison of bulk hBN PDOS with the VPS of a $NVE$ MD simulation. In panel b), we compare the VPS of the $NVE$ simulation (broadened by a Lorentzian of 1.2~meV FWHM) with the VPS of a $NVT$ simulations using a Langevin thermostat with damping parameters $\tau$ of 0.5~ps and 0.05~ps. In panel c) we compare the $NVE$ VPS (broadened by a Lorentzian of 12~meV FWHM) and the VPS obtained from a simulation with Langevin thermostat with damping 0.05~ps. Note that the broadened VPS and the VPS from a Langevin thermostat calculation with $\tau=0.5$~ps are practically indistinguishable in panel b). All spectra are normalized to unit integral between 0 and 220~meV. Arbitrary units (a.u.).}
    \label{fig:MD_PDOS_hBN}
\end{figure}

\subsection{Influence of simulation box size on the MD simulation\label{sec:finite_size}}
In order to simulate bulk properties, we need to introduce periodic boundary conditions on a finite size simulation box in the MD simulations. This leads to a certain dependence of computed properties on the size of the simulation box. For our method it is of interest, that periodic boundary conditions limit the density of the grid of possible wave vectors of phonons, which can be present in the simulation: the smaller the simulation box, the fewer wave vectors can satisfy the boundary conditions. In extreme cases, the VPS of a small simulation box can appear as a set of discrete peaks -- especially for long MD trajectories, which allow a finer step on the energy axis of the VPS. Such a system would clearly not be a good representations of bulk hBN in the sense of reproducing correct phonon properties as stated earlier. It should be noted here, that this issue is not unique to MD simulations: in a quantum mechanical approach to vibrational EELS simulations, one needs to compute transition matrix elements associated with a finite set of phonon modes. Each matrix element depends explicitly on the displacements of atoms associated with its mode and one thus encounters therefore similar finite size effects. For large systems and sufficiently low temperatures, such that anharmonic effects are largely suppressed, the VPS of the MD model system converges gradually towards the PDOS of the bulk. Our first step is thus to compare the VPS of our structure model to the bulk PDOS in order to asses how influential finite box size effects are in our simulation.

We show in Fig.~\ref{fig:MD_PDOS_hBN}a) a comparison of the bulk PDOS with a unbroadened and a broadened VPS computed from a $NVE$ MD simulation at average kinetic energy corresponding to a temperature of around 300~K. The broadening function is a Lorentzian of full width at half maximum (FWHM) of 1.2~meV. While unbroadened VPS and bulk PDOS have a similar shape as a function of energy, the VPS is much noisier with some large spikes and a few small gaps in the spectrum. These gaps and spikes are a manifestation of finite size effects in our system. The broadened VPS fluctuates around the bulk PDOS for the largest part, but all of the main peaks of the PDOS are reproduced by the broadened VPS, albeit we observe a slight red-shift of the main optical peak around 190~meV in both the unbroadened and broadened VPS, which is likely a consequence of anharmonicity of the potential. Consistent with this interpretation, separate simulations at larger average kinetic energies, which are not shown here, showed an even larger shift of the optical peak towards lower frequencies.

Overall, since the mildly broadened VPS follows closely the PDOS, we conclude that although finite size effects are present, they are rather weak and we do not miss any essential features of the phonon dispersion. Moreover, experiments necessarily introduce additional broadening due to finite energy resolution, blurring the remaining differences even more. Yet, in Sec.~\ref{sec:freq_therm} below we will illustrate that under certain circumstances the finite size effects may still play a role in FRFPMS EELS simulations.

\subsection{Influence of the thermostat}
As discussed in the methods part, thermostats modify the dynamics of atoms in MD simulations. Therefore we also need to take the action of the thermostat into account when comparing the VPSs to each other and to the PDOS. In Fig.~\ref{fig:MD_PDOS_hBN}b) we show how the VPS changes by applying a Langevin thermostat on top of the $NVE$ dynamics. We have considered two different settings for the damping parameter $\tau$, namely 0.5~ps and 0.05~ps. We remind the reader that the actual friction force is inversely proportional to $\tau$, i.e., larger damping parameter leads to smaller friction and noise terms, see Eq.~\ref{eq:Langevin_equation}. Both settings for the thermostat introduce broadening into the VPS. As could be expected, the thermostat with a larger value of the damping parameter leads to a smaller broadening, which is equivalent to a broadening of the pure $NVE$ VPS by a Lorentzian of FWHM of 1.2~meV. A damping parameter of 0.05~ps smears the VPS significantly and most of the noisiness of the original VPS has disappeared. We show in Fig.~\ref{fig:MD_PDOS_hBN}c) that the spectral shape is roughly equivalent to the $NVE$ VPS broadened by a Lorentzian of 12~meV FWHM. At the same time, the height of the TO peak at 46~THz has decreased considerably and the lower frequency peaks have acquired a very broad shape. 

Thus we can conclude, that applying a Langevin thermostat leads to a damping parameter dependent broadening of the $NVE$ VPS, which can be approximately modeled by a Lorentzian. We note furthermore, that the broadening also reduces the spikiness due to finite size effects at the cost of resolution. In the following paragraphs, we will show how the VPS is influenced by the frequency-dependent thermostats and compare their VPSs to the broadened VPSs shown in Figure~\ref{fig:MD_PDOS_hBN}.

\begin{figure}
    \centering
    \includegraphics[width = \linewidth]{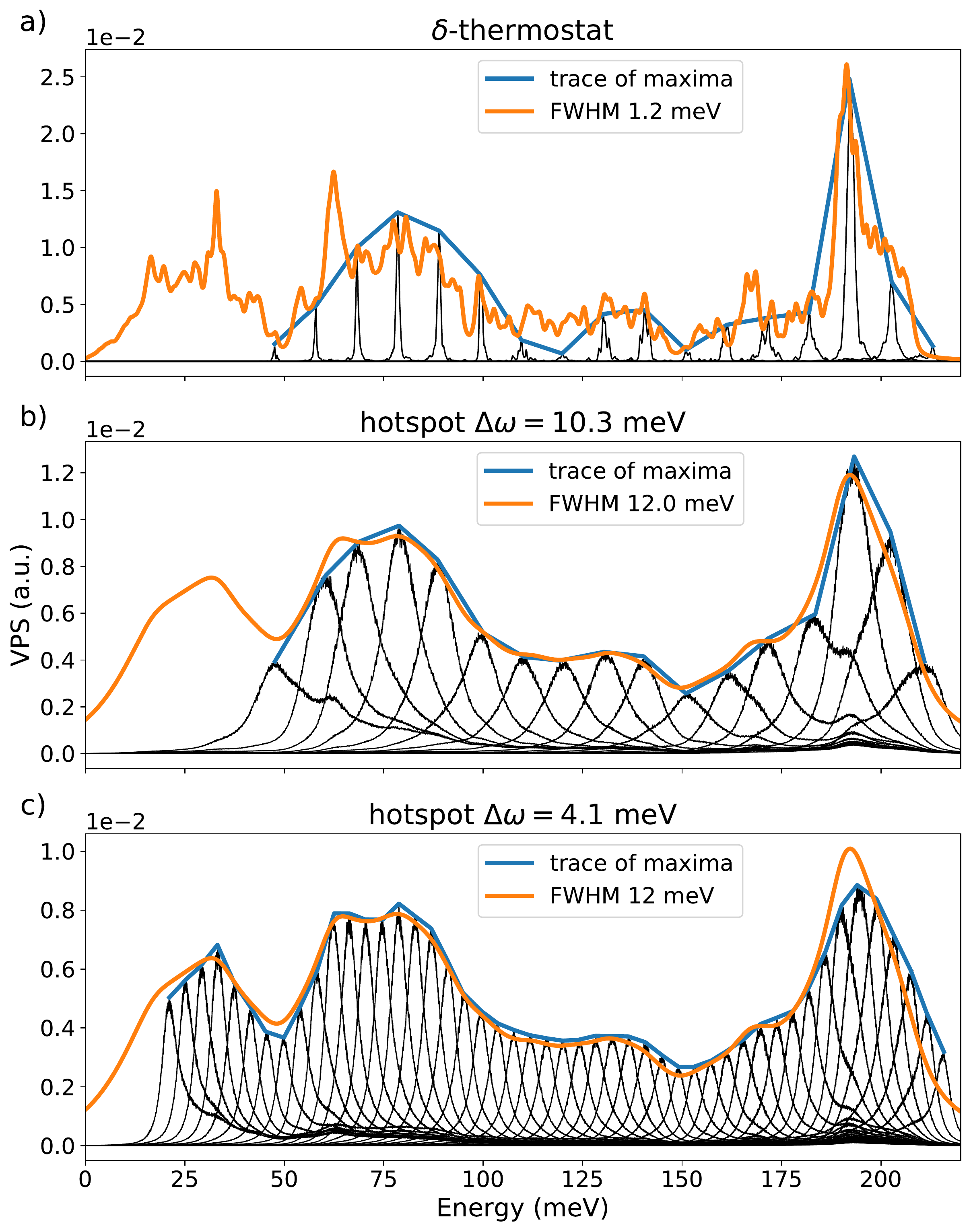}
    \caption{Comparison of the VPSs of $\delta$-thermostat and hotspot thermostats with the broadened $NVE$ VPS. The black curves correspond to the VPSs for the different energy bins. The broadened $NVE$ VPSs  have already been shown in Fig.~\ref{fig:MD_PDOS_hBN}. The ``trace of maxima'' is a guide for the eye to facilitate comparison with the reference VPS and indicates the maximum value of the VPS for each energy bin. The overall scaling of individual VPSs of $\delta$- and hotspot thermostats is such that the integral of their trace of maxima is one between 48 and 215~meV in panel a) and b), and 20 and 215~meV in panel c). Arbitrary units (a.u.).}
    \label{fig:thermostat_comparison}
\end{figure}

\subsection{VPSs of frequency-dependent thermostats\label{sec:freq_therm}}

We have seen in the previous paragraph, how the commonly used Langevin thermostat influences the VPS of the MD trajectory. This immediately leads to the question, how the $\delta$-thermostat and hotspot thermostats used in the FRFPMS method influence the VPS and how the spectrum of vibrations excited by the thermostats actually looks like. Since the VPS considers the range of vibrations, which is present in the corresponding MD trajectories, the energy resolution of the FRFPMS method cannot be better than the resolution observed in the VPS. Therefore, we compare in Fig.~\ref{fig:thermostat_comparison} plots of the VPSs as a function of energy for these thermostats and their ``trace of maxima'' with the broadened $NVE$ VPS. Generally the magnitude of the peaks of the VPS follows the broadened $NVE$ VPS in all energy bins and for all thermostats. In the details, we observe, however, some important differences between the thermostats.

The $\delta$-thermostat in Fig.~\ref{fig:thermostat_comparison}a) excites vibrational energies within a very narrow interval. For our considered grid of $\delta$-thermostats, this results in a series of non-overlapping VPS peaks. For most energies, the magnitude of these peaks correlates well with the mildly broadened $NVE$ VPS. At around 120~meV, we observe, however, an unexpectedly small peak, which does not correlate well with the broadened $NVE$ VPS. This highlights an important issues of the $\delta$-thermostat for FRFPMS simulations: due to its high frequency selectivity, it is highly sensitive to the exact details of the raw unbroadened VPS. In theory this can be considered an advantage, as it would allow for a high energy resolution. In practice, however, finite size effects, which we have discussed in Sec.~\ref{sec:finite_size}, can have a significant influence on the observed spectra, since the spectral shape can vary quite significantly as a result of slight shifts in the positioning of the energy bins. One could easily imagine a case, where a significant, but narrow spike in the VPS is missed entirely by $\delta$-thermostat calculations or that it exaggerates small differences, unless a very dense grid of such thermostats is used. It is in this context, where the spikiness of unbroadened VPS due to finite size effects can matter.

As was mentioned, the issue could be mitigated by having a very dense grid of $\delta$-thermostats and use an appropriate smearing of the spectra, or by using a larger structure model, which would smooth the VPS and bring it closer to the bulk PDOS. Both procedures would encompass increasing computational costs, potentially to an impractical level. On the other hand, the results in panels b) and c) of Fig.~\ref{fig:thermostat_comparison} show, that the VPSs of individual simulations with a hotspot thermostat compare nicely to a broadened $NVE$ VPS, which shows much fewer signs of finite size effects. This suggests, that switching out the $\delta$-thermostat for a hotspot thermostat, which excites vibrations within a larger energy window, should lead to a more predictable spectral shape. With our choice of thermostat parameters, this comes at the cost of lower energy resolution as panels b) and c) of Fig.~\ref{fig:thermostat_comparison} show.

At this point we should mention our initial FRFPMS report \cite{zeiger_efficient_2020}. There we have not studied the actual VPS and as such the calculated spectral shapes could be subject to artifacts due to finite size effects. But even then, sums over several energy bins, such as those performed in calculation of atomic resolution images, would only be impacted to a smaller extent -- as is supported by their good comparison to earlier results \cite{hage_phonon_2019}.

We turn our attention now towards considering the differences between the hotspot thermostat with width parameter $\Delta \omega$ of 10.3~meV (2.5~THz) and 4.1~meV (1~THz), respectively. We observe in line with expectations, that a smaller width parameter indeed leads to a smaller width of the peak in the corresponding hotspot VPS. However, the energy resolution of both settings is very similar as indicated by the ``trace of maxima''. Furthermore the main optical peak appears to be shifted for a hotspot thermostat with width parameter of 4.1~meV. Such a shift is not observed for a hotspot thermostat with width parameter of 10.3~meV. The shift could be due to the chosen grid of the peak energies, but we have not further investigated this circumstance. Lastly, we observe that the peaks of the VPSs of the hotspot thermostat have rather large tails in comparison with the delta thermostat.

Considering the VPSs of the hotspot thermostat with a width of 4.1~meV displayed in Fig.~\ref{fig:thermostat_comparison}c) more closely, we note, that the ``trace of maxima'' nicely reproduces the double peak structure at about 65 and 80~meV as well as the small peak around 170~meV on the low energy shoulder of the main optical peak. The broadened $NVE$ VPS agrees furthermore best with the ``trace of maxima'' for intermediate energies between around 55 to 185~meV. Outside this interval, we observe larger deviations. A detailed consideration of different values for the FWHM of the Lorentzian broadening, which is not shown here, revealed, that a FWHM of 10~meV fits better to the ``trace of maxima'' at lower energies, whereas at higher energies a FWHM of 16~meV yields a better fit. %

We can conclude from this paragraph, that we resolve issues connected to the finite structure model with the hotspot thermostat at the cost of lowering the attainable energy resolution in the calculation, but keeping computational costs unchanged. The hotspot thermostat leads to dynamics, which exhibit a clear peak at the desired vibrational energy, but the peak has rather large tails. Additionally the width of the hotspot thermostat depends somewhat on the peak energy and the chosen width parameter, but the attainable energy resolution is largely unaffected by the width parameter alone. These observations may give hints how to optimize the colored thermostats further for the FRFPMS method.

\begin{figure}
    \centering
    \includegraphics[width = \linewidth]{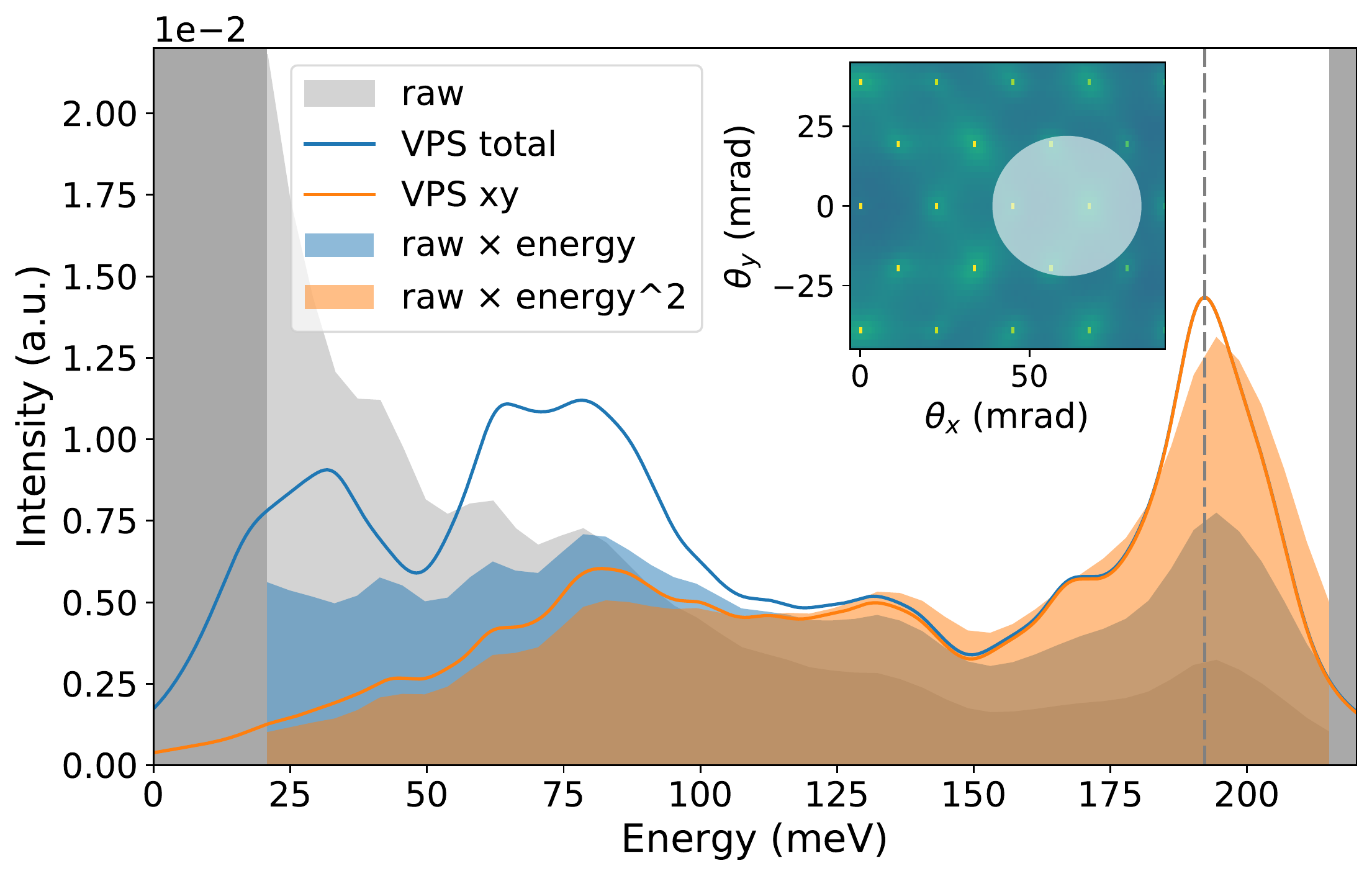}
    \caption{Comparison of off-axis EELS spectrum with the VPS. The inset shows how the center of the off-axis detector is displaced by 61~mrad along the $\theta_x$-direction (parallel to $\Gamma$-M line) in the diffraction plane in this off-axis geometry. The dark grey areas marks the energy range, which is not covered by FRFPMS calculation. Total VPS and raw spectrum normalized to unit integral between energies of about 40 and 215~meV. The $x/y$-projected VPS is normalized such that it matches the peak at 190~meV of the total VPS. Arbitrary units (a.u.).}
    \label{fig:cmpPDOS-raw-energmult}
\end{figure}

\subsection{Comparison of off-axis, large collection angle vibrational spectra with the VPS}

We turn now to the FRFPMS simulations. In the first step, we ask ourselves, in which way the FRFPMS EELS signal encodes the VPS. To that end, we compare in Fig.~\ref{fig:cmpPDOS-raw-energmult} an off-axis spectrum for a large collection angle with the VPS. We have used the detector geometry reported in Hage et al. \cite{hage_phonon_2019}, i.e., a collection semi-angle of 22~mrad with the detector displaced from the center of the diffraction pattern by 61~mrad along the $\Gamma$-M line. Note that such a detector covers 4 $\Gamma$-points (Bragg reflections, c.f. the inset of Fig.~\ref{fig:cmpPDOS-raw-energmult}), where convergence is difficult to achieve as mentioned in section~\ref{sec:methods_MD}. Therefore it might be adequate considering to exclude these points from detector integration in Fig.~\ref{fig:cmpPDOS-raw-energmult}. However, the spectra at $\Gamma$-points are smooth and mainly contribute a steep and rapidly decreasing, yet featureless background to the large collection angle off-axis spectra (not shown here), which does not qualitatively change the overall spectral shape. Therefore, and for the sake of simplicity, we did not remove the $\Gamma$-points from the detector integration in Fig.~\ref{fig:cmpPDOS-raw-energmult}.

The raw FRFPMS vibrational EELS (gray-shaded area) decreases rapidly as a function of energy. Several peaks can be observed, the most pronounced ones are located around 40, 65, 80, 140, and 190~meV. Inelastic phonon excitation matrix elements have an explicit dependence on $1/\omega$, where $\omega$ is the angular frequency (energy) of the phonon mode on which the electron is scattered \cite{senga_position_2019,hage_single-atom_2020}. In order to cancel out this global $1/\omega$-dependence and to reveal more of the structure of the spectrum, we multiply the raw spectrum by energy. Furthermore we also show the spectrum multiplied by the energy squared. Both resulting spectra show also peak structures at 40, 65, 80, 140 and 190~meV.

Starting from low energies in Fig.~\ref{fig:cmpPDOS-raw-energmult}, the broadened total VPS exhibits a broad peak around 30~meV, followed by a relatively large dip around 50~meV and a broad double peak structure between 50 to 100~meV. Towards larger energies, another strong dip is observed at 150~meV before the shoulder of the main optical peak, which is located around 190~meV. On the left shoulder of that peak, we discern a small plateau around 170~meV. The broadened $x/y$-projected VPS is practically equal to the broadened total VPS for energies above around 130~meV. Below this energy, the $x/y$-projected VPS has a significantly lower magnitude than the total VPS. It furthermore exhibits a peak around 80~meV and two minor peaks around 63 and 45~meV. The VPS and $x/y$-projected VPS compare well to the PDOS and $x/y$-projected PDOS displayed in Fig.~\ref{fig:Figure_banddos}.

We will turn now towards the question, how well the raw off-axis EELS agrees with the total VPS and the $x/y$-projected VPS. First we compare peak positions and starting at large energies, we recognize, that the maximum value of the main optical peak around 190~meV appears in EELS at a position, which is very similar to the position of the corresponding peak in the VPS. A slight shift can be discerned, which stems likely from the circumstance, that the exact peak energy in the VPS lies between two points of our grid of hotspot thermostats. The small plateau structure in the VPS around 170~meV is smeared into a flat shoulder and the dip at 150~meV appears at a slight shift in the EELS with respect to the VPS. Both VPS and EELS exhibit a small peak at 130~meV. At lower energies, we note that there are two minor bumps in the $x/y$-projected VPS as well as the EELS around 100 and 120~meV. The positions of peaks at 65 and 80~meV agree furthermore well with the position of the peaks in the broad double peak structure of the total VPS. Similarly, these two peaks correlate with the step structure in the $x/y$-projected VPS. However, the peak visible at 40~meV in the EELS spectrum compares better with a similar feature in $x/y$-projected VPS than with the total VPS in this energy range.

We focus our attention now on the shape of the raw and energy-multiplied spectra and how the total shape compares to the shape of the VPS. It is clear, that the raw EELS spectrum contains some overall energy-dependent scaling, since its shape, apart from the peak positions, agrees very little with the VPS. The EELS multiplied by energy shows a good agreement with both the total as well as the $x/y$-projected VPS at energies above about 130~meV. Only the relative size of the main optical peak around 190~meV is somewhat smaller. At lower energies, the energy-multiplied spectrum does not exhibit the same dip around 50~meV as the total VPS and it does not decrease as rapidly as the $x/y$-projected VPS, yet all its peak positions align with the $x/y$-projected VPS.

Lastly we observe a very good agreement of the shape of the large collection angle off-axis EELS multiplied by the square of the energy and the $x/y$-projected VPS. In the vibrational scattering process, the momentum transfer in $z$-direction (out-of-plane) should be very small compared to the momentum transfer in the $x/y$-plane (in-plane). Therefore the the EELS signal should be dominated by scattering on vibrational modes, whose $\mathbf{k}$-vector lies in the $x/y$-plane. In this sense it is not surprising, that we see good agreement with the $x/y$-projected VPS. It is, however, somewhat surprising, that the best agreement is found after multiplying the raw EELS by the square of the energy, rather than by energy only, as is the case in the Born approximation treatment. For the moment, we do not have a qualitative explanation how such an energy scaling could arise. Removing $\Gamma$-points from the detector integration reduces the background of the low-energy region of the large collection angle off-axis spectra by 5--10\% (not shown here), but this measure is not enough to substantially influence the overall energy scaling. Therefore, the issue can not be connected with convergence-related concerns at $\Gamma$-points alone. The most likely hypothesis is, that the tails of the VPSs of the hotspot thermostat together with the $1/\omega$-divergence of the EELS cross section at low energies could introduce such energy-dependent scaling in the spectra. To illuminate this point a bit further, consider the relative magnitude of the tails of the VPSs corresponding to two energy bins near zero energy. The VPS corresponding to the lower energy bin will exhibit a larger value near zero, than the VPS corresponding to the higher energy bin. The difference will be small, but it could prove enough to introduce a background, which leads to the observed energy scaling. Reducing the zero-energy tails of the excited energy windows could thus lead to an energy scaling, which would be closer to $1/\omega$.

Overall, we can state that the positions of peaks agree very well particularly between the $x/y$-projected VPS and the FRFPMS large detector off-axis EELS. The question of the nature of the energy-scaling of the FRFPMS EELS remains a subject for further investigations, which is likely connected to further refined thermostat schemes.

\begin{figure}
    \centering
    \includegraphics[width = \linewidth]{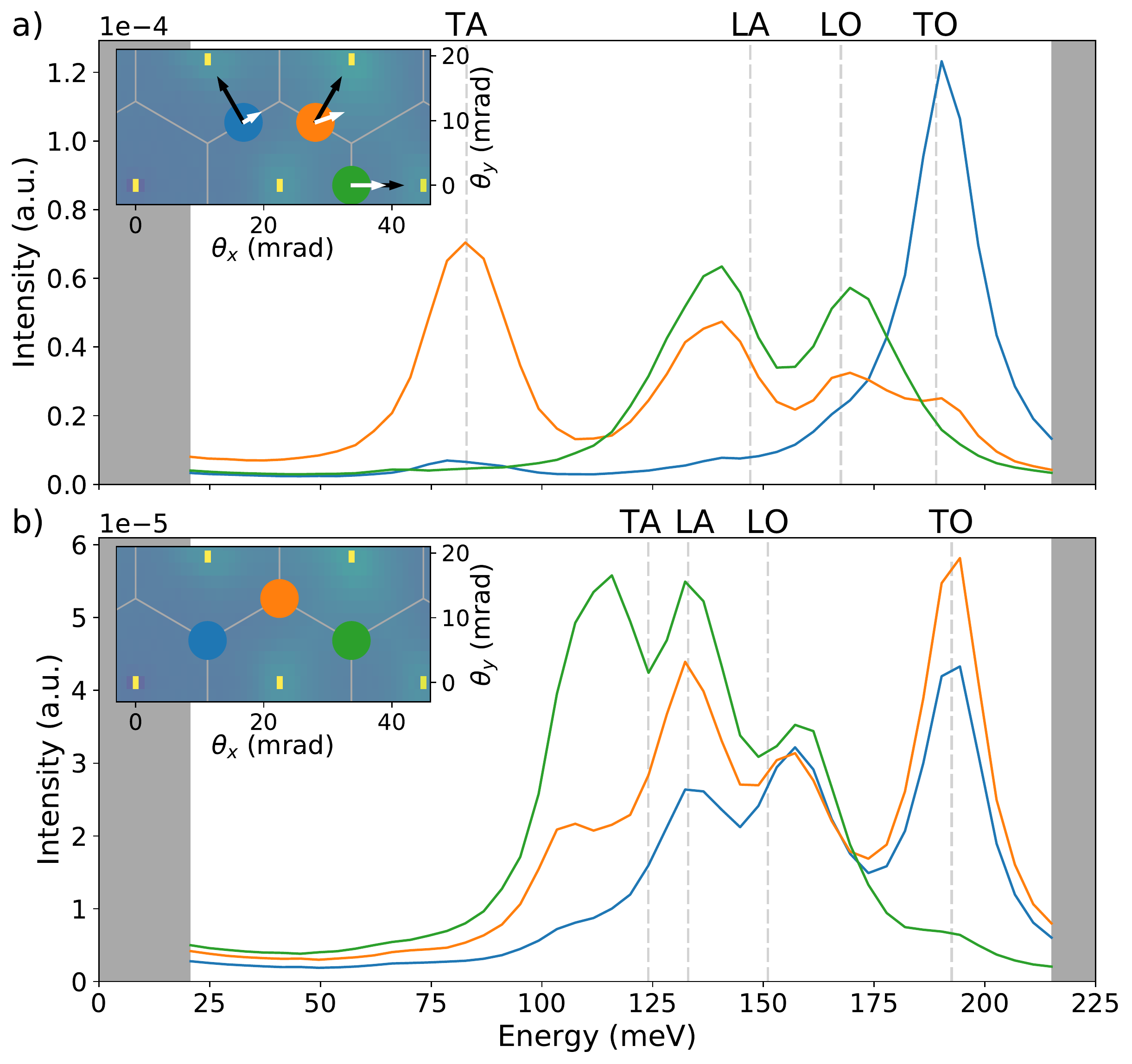}
    \caption{Spectral variations in FRFPMS EELS as a function of detector position in the diffraction plane at a thickness of about 2.5~nm. Panels a) and b) show a comparison of energy-multiplied EELS spectra for a detector entrance aperture centered around M- and K-points, respectively, in the diffraction plane. The collection semi-angle of the detector is in all cases 3~mrad. The light grey dashed lines indicate the positions of the modes indicated at the top of the figure, which we expect from phonon bandstructure, see Fig.~\ref{fig:Figure_banddos}. The inset shows the detector position in the diffraction plane corresponding to the spectrum of the same colour. The black arrows in the inset of panel a) indicate the direction of the phonon polarization vectors of the longitudinal phonon modes at the M-point and white arrows indicate the direction of the momentum transfer. Lightgrey hexagons indicate boundaries of reciprocal space Wigner-Seitz unit cells.}
    \label{fig:cmp_spec_kpoints_collang3e+00mrad_thickness125}
\end{figure}

\begin{figure}
    \centering
    \includegraphics[width = \linewidth]{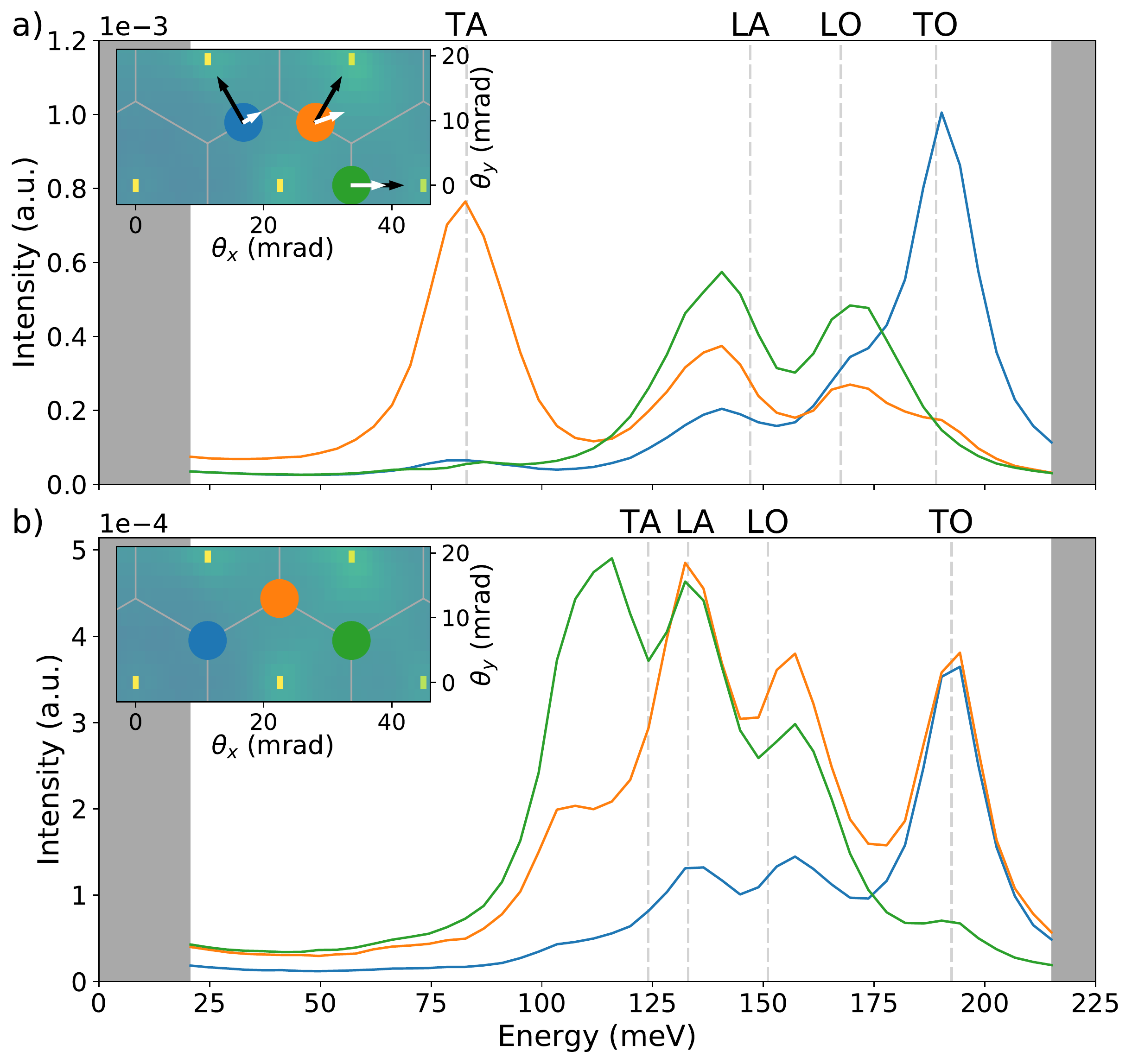}
    \caption{The same as  Fig.~\ref{fig:cmp_spec_kpoints_collang3e+00mrad_thickness125} for sample thickness of 15~nm.}
    \label{fig:cmp_spec_kpoints_collang3e+00mrad_thickness750}
\end{figure}

\subsection{Spectral changes as a function of detector position and thickness}
\label{sec:results_spectral_change_as_function_of_detector_position_thickness}

We focus in this paragraph on qualitative changes in the spectra as we move a detector to a set of three points in the diffraction plane, which can be mapped onto the same point within the irreducible wedge of the Brillouin zone. Figures~\ref{fig:cmp_spec_kpoints_collang3e+00mrad_thickness125} and \ref{fig:cmp_spec_kpoints_collang3e+00mrad_thickness750} show comparisons of such spectra collected at various K- and M-points at a crystal thickness of 2.5 and 15~nm, respectively. Experimentally it is feasible to obtain spectra under similar conditions as for example Hage et al., Nicholls et al., and Senga et al.\ have reported \cite{hage_nanoscale_2018,nicholls_theory_2019,senga_position_2019}. We note furthermore, that in the Born approximation treatment, the vibrational EELS cross section depends on the square of the absolute value of the scalar product $\mathbf{e}\cdot\mathbf{q}$ of the phonon polarization vector $\mathbf{e}$ with the momentum transfer $\mathbf{q}$. Exemplified by the spectra at the M-point, we will see in the following, that the EELS calculated by the FRFPMS method changes consistently with such dependence. Furthermore non-trivial thickness effects are also present in the FRFPMS spectra, likely due to dynamical diffraction. Note that the possibility to include such effects is one of the major strengths of the FRFPMS method, since theoretical treatments based on the Born approximation do not allow for the inclusion of probe deforming elastic diffraction before or after the inelastic scattering event.

As a first observation from Figs.~\ref{fig:cmp_spec_kpoints_collang3e+00mrad_thickness125} and \ref{fig:cmp_spec_kpoints_collang3e+00mrad_thickness750}, we note that none of the spectra show peaks around 60~meV or lower. Comparing with the phonon dispersion in Fig.~\ref{fig:Figure_banddos} leaves us thus to conclude, that vibrations in $z$-direction, i.e. out-of-plane vibrations, do not contribute noticeably to the FRFPMS EELS in the present calculation.

We focus in panel a) of  Figs.~\ref{fig:cmp_spec_kpoints_collang3e+00mrad_thickness125} and \ref{fig:cmp_spec_kpoints_collang3e+00mrad_thickness750} on the selected M-points in the diffraction plane. At the M-point the direction of the phonon polarization vectors $\mathbf{e}$ of different phonon modes can be easily inferred from the name of the corresponding mode, i.e., longitudinal modes have phonon polarization vectors parallel to the wave vector and the phonon polarization vectors of transversal modes point perpendicular to the corresponding phonon propagation direction. A nice visualisation of the phonon modes of different materials, among them hBN, is available on the website of Ref.~\cite{phonon_visualization_web}. From the phonon dispersion in Fig.~\ref{fig:Figure_banddos}, we expect to observe peaks corresponding to the different phonon modes in the EELS spectrum at positions indicated by the grey dashed lines.

We will discuss in the following the spectra corresponding to every choice of detector position in panels a) of Figs.~\ref{fig:cmp_spec_kpoints_collang3e+00mrad_thickness125} and \ref{fig:cmp_spec_kpoints_collang3e+00mrad_thickness750}, which correspond to a detector centered around the M-point in the diffraction pattern. We note first, that we observe peaks in good agreement with the expected positions of TA, LO and TO modes in the FRFPMS EELS, but the peaks corresponding to the LA mode is shifted towards lower energy. We attribute this shift in the peak position to the finite size of the detector collection angle of 3~mrad, which effectively averages the spectra corresponding to several different momentum transfers. The TA and LO modes are not affected by this circumstance at the M-point, since the dispersion shows here a saddle point, which means that in some vicinity around the M-point, the corresponding mode has lower as well as higher energies than at the M-point. Also the TO mode is not strongly influenced by the finite detector aperture, since its dispersion is quite flat around the M-point. The LA mode on the other hand shows a local maximum at the M-point, which means, that the LA mode has a lower energy at all points in a close vicinity around the M-point. Consistent with this observation, spectra integrated over a smaller collection angle show peak positions much closer to the positions expected from the phonon dispersion (not shown).

For both thicknesses, the TO mode dominates the blue colored spectrum. The LO mode is barely discernable from the background at a thickness of 2.5~nm, whereas it appears as shoulder of the TO at a thickness of 15~nm. The peak corresponding to the LA mode is barely visible at low thickness, but appears more strongly at larger thickness. The TA mode gives rise to a weak signal and its relative strength is similar for both thicknesses. 

In the orange spectrum on the other hand all modes are visible to some degree at both thicknesses. We observe furthermore strong signal corresponding to the TA mode, but also LA and LO mode are visible as peaks. The TO peak is much reduced in comparison with the blue spectrum and appears only as a shoulder of the LO peak. Thickness effects manifest themselves mainly in the relative weights of peaks corresponding to individual modes.

Finally, the green spectrum exhibits strong signal for LA and LO modes, but shows no hint of TA and TO modes at both thicknesses. The shape of the green spectrum is furthermore almost unaffected by thickness effects.

If we concentrate on the longitudinal modes around 145-150~meV and 165-170~meV, we notice, that their relative strength in the spectrum is lowest at the blue detector position and a bit larger at the orange position. At the green detector position, the LA and LO modes make up almost the entire spectrum. This observation is consistent with a dependence of the spectrum on the scalar product $\mathbf{e}\cdot\mathbf{q}$ described earlier.

We observe furthermore, that the signal corresponding to the TA mode is extremely weak in the blue spectrum. If the visibility of peaks depends mainly on the scalar product $\mathbf{e}\cdot\mathbf{q}$, we would expect the EELS corresponding to the TA mode to be stronger in the blue spectrum than in the orange one, since the transverse mode maximizes the scalar product. At the moment we do not have a simple qualitative picture to explain this observation. The corresponding peak is visible but relatively weak in Fig.~10 of Ref.~\cite{plotkin-swing_hybrid_2020}, suggesting a possibility of some dynamical mechanisms leading to a destructive interference.

Moving on to the spectra around the K-points in panels b) of Figs.~\ref{fig:cmp_spec_kpoints_collang3e+00mrad_thickness125} and \ref{fig:cmp_spec_kpoints_collang3e+00mrad_thickness750}, we see a similar shift of peaks away from their expected positions as in panels a) of the same figures. Here it is however the TA and LO modes, which exhibit local maxima and minima, respectively, as displayed in Fig.~\ref{fig:Figure_banddos} and consistent with the observed deviations. TO and LA modes are not shifted significantly because their dispersion is either relatively flat near the K-point (TO mode) or exhibits a saddle point (LA mode).

Starting with the blue spectrum in panels b) of Fig.~\ref{fig:cmp_spec_kpoints_collang3e+00mrad_thickness125} and \ref{fig:cmp_spec_kpoints_collang3e+00mrad_thickness750}, the spectrum exhibits strong peaks corresponding to LA, LO and TO modes, but very weak signal corresponding to the TA mode, which is also shifted towards lower energies by about 20meV. Thickness effects manifest themselves mainly in the relative weights of LA and LO peaks to the TO peak. The signal corresponding to the TA mode has a similar relative strength for both thicknesses.

\begin{figure}
    \centering
    \includegraphics[width = \linewidth]{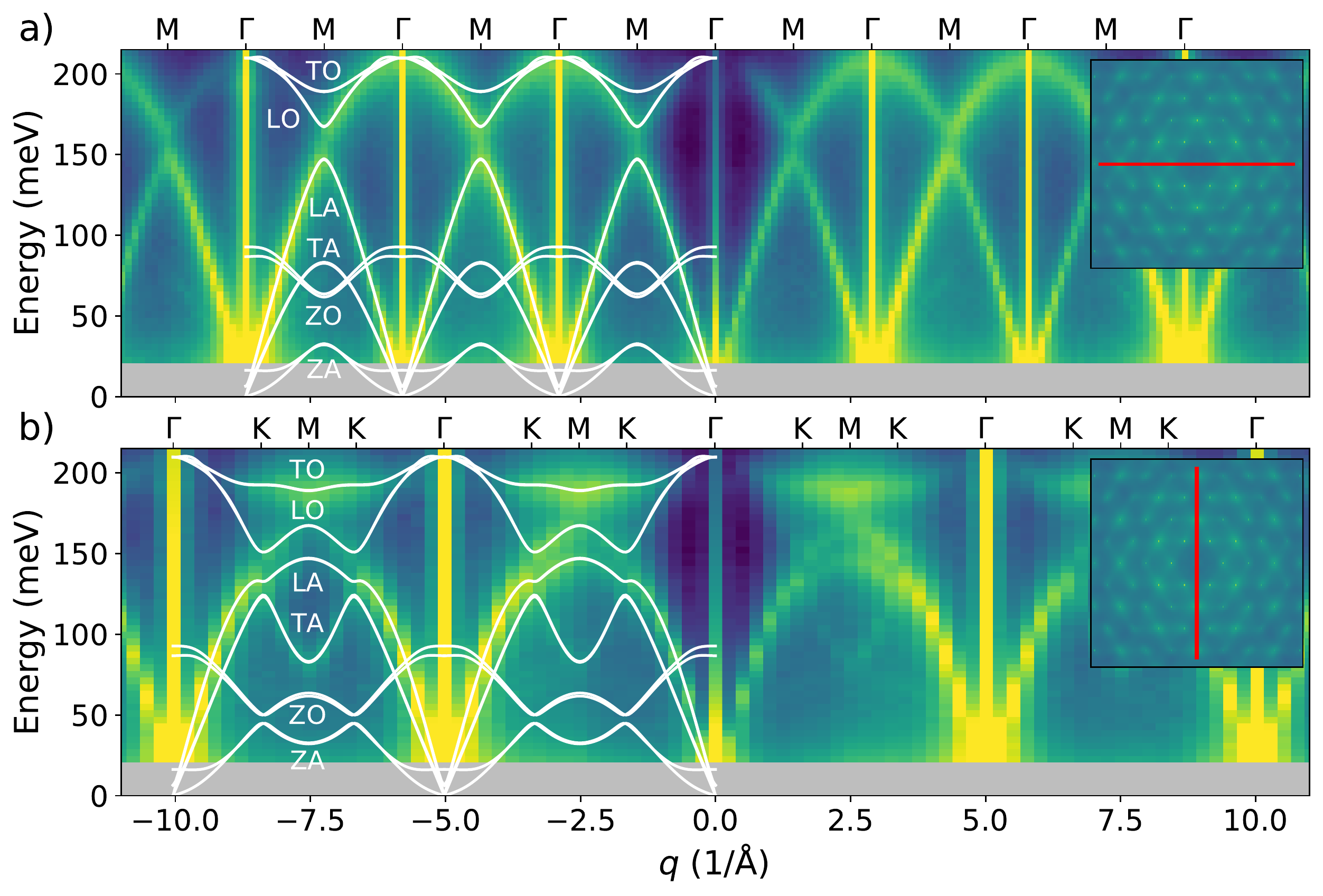}
    \caption{Comparison of the raw vibrational EELS signal along selected directions in the diffraction pattern with the phonon dispersion relation (white lines). The directions lie in a) and b) along the $\Gamma$-M-$\Gamma'$ and $\Gamma$-K-M-K-$\Gamma'$ line, respectively. Both paths are indicated relative to the diffraction pattern by the red line in the insets. The logarithmic color scale ranges from dark blue for small values of the EELS signal over blue, turquoise and green to yellow for large values. The EELS signal was not calculated for energies corresponding to the grey area.}
    \label{fig:cmp_TDS_phonon_disp}
\end{figure}

The orange spectrum exhibits strong features of all phonon modes, whereby the TA mode appears as a shoulder of the LA peak. Thickness effects affect mainly the relative weight of the TO peak to the rest of the spectrum. At lower thickness, the relative weight of the TO peak is higher than for thicker sample. The relative weights of the peaks in the green spectrum do not change much with thickness. However the shift of the peak corresponding to the TA mode is about 10~meV compared with the energy of the TA mode at the K-point in the green spectrum, whereas it is about 20~meV in the blue and orange spectra. This could be a consequence of direction dependent weigthing of different modes at different points in the diffraction plane together with an asymmetry in the steepness of the TA mode at the K-point (c.f. Fig.~\ref{fig:cmp_TDS_phonon_disp}).

At the K-point the phonon polarization vectors are complex valued, which means, that the atoms perform vibrations along circular or elliptical trajectories and that the scalar product $\mathbf{e}\cdot\mathbf{q}$ becomes complex-valued. Such complex-valued scalar product is less selective, than the real scalar product encountered at the M-point, since it can not be ``zeroed'' by the real momentum transfer $\mathbf{q}$ and we do not observe a similarly strong suppression of modes as at the M-point. Consequently, all spectra show signs of all modes, but a comparison of the relative peak intensities between spectra reveals a non-trivial directional and positional scaling of individual peaks in the diffraction plane at the K-point.

One could study at this point many of such spectra and compare them to PDOSs integrated over small areas in reciprocal space instead of the full Brillouin zone. We chose in the following section a different route in order to present the results in a more compact, yet still experimentally accessible way: we compare the phonon dispersion along certain directions in reciprocal space with the TDS signal along the corresponding directions in the diffraction plane. Using the latest generation of highly efficient detectors, the necessary measurements of spectra at different positions in the diffraction pattern can be performed in parallel as showcased by Plotkin-Swing et al. \cite{plotkin-swing_hybrid_2020}.

\subsection{Comparison of the FRFPMS EELS with the phonon dispersion}

In Fig.~\ref{fig:cmp_TDS_phonon_disp} we compare the vibrational EELS with the phonon dispersion along the $\Gamma$-M-$\Gamma$ and $\Gamma$-K-M-K-$\Gamma$ reciprocal space paths in the diffraction plane. Generally, we observe the formation of band-like structures in the EELS signal. The EELS intensity is symmetric with respect to zero momentum transfer, but the intensity of the bands depends on which reciprocal unit cell one is considering. For example, in Fig.~\ref{fig:cmp_TDS_phonon_disp}a) the LA mode appears at medium intensity in the EELS signal for momentum transfers between 0~{\AA}$^{-1}$ and 1.45~{\AA}$^{-1}$. In the next reciprocal unit cell, the EELS intensity stays first at a medium level, but becomes then stronger towards the $\Gamma$-point located at a momentum transfer around 2.9~{\AA}$^{-1}$. Between 2.9~{\AA}$^{-1}$ and 4.35~{\AA}$^{-1}$ the EELS signal corresponding to the LA mode reaches a large value, before becoming smaller again between 4.35~{\AA}$^{-1}$ and 5.8~{\AA}$^{-1}$. This pattern is also observed in the simulations and experiments of Senga et al. \cite{senga_position_2019}. Similar intensity variations can be observed in the EELS signal corresponding to other branches of the phonon dispersion.

At momentum transfers corresponding to the $\Gamma$-point in reciprocal space, i.e., at momentum transfers of 0, $\pm2.9$, $\pm5.8$~{\AA}$^{-1}$, $\pm8.7$~{\AA}$^{-1}$, and 0, $\pm5$, $\pm10$~{\AA}$^{-1}$ along $\Gamma$-M-$\Gamma$ and $\Gamma$-K-M-K-$\Gamma$, respectively, we observe that the vibrational EELS signal is much stronger at almost all energies than at surrounding points in the diffraction plane. The signal decreases furthermore smoothly and quickly as a function of energy at all $\Gamma$-points. Deeper spectral investigations in the $\Gamma$-M-$\Gamma$ direction showed, that the spectra at the central $\Gamma$-point and at the $\Gamma$-point around $\pm2.9$~{\AA}$^{-1}$ do not contain any peaks, which could be attributed to phonon modes (not shown here). This observation is puzzling given that the LO/TO mode is visible at surrounding points in the diffraction plane. Spectra at the $\Gamma$-points around $\pm5.8$~{\AA}$^{-1}$ on the other hand exhibit a small peak around 210~meV, whereas the spectra at the $\Gamma$-points at $\pm8.7$~{\AA}$^{-1}$ again do not exhibit such signal. Plotkin-Swing et al., observe similarly a relatively strong phonon signal at $\Gamma$-points \cite{plotkin-swing_hybrid_2020}, whereas Senga et al. do not observe a strong vibrational EELS for all energy losses at $\Gamma$ \cite{senga_position_2019}, neither in experiment nor in theory. Nevertheless, we remind the above-mentioned, potential convergence issues in these points of reciprocal space. Therefore our observations within this paragraph should be taken cautiously.

We turn now towards features of Fig.~\ref{fig:cmp_TDS_phonon_disp} in between the $\Gamma$ points, where no convergence issues are present. Considering each branch of the phonon dispersion in more depth, we recognize, that the LA mode is well represented along both directions in the diffraction pattern. The LA branch exhibits furthermore a modulation in the EELS intensity of the LA mode along the different $\Gamma$-M segments in Fig.~\ref{fig:cmp_TDS_phonon_disp}a). These observations agree with other published data \cite{senga_position_2019,nicholls_theory_2019,plotkin-swing_hybrid_2020}.

The TA mode is essentially absent along $\Gamma$-M-$\Gamma$ in Fig~\ref{fig:cmp_TDS_phonon_disp}a), consistent with Born approximation, due to the zero scalar product of $\mathbf{e \cdot q}$. TA it appears along the path $\Gamma$-K-M-K-$\Gamma$ in Fig~\ref{fig:cmp_TDS_phonon_disp}b) between the $K$-points, albeit its intensity is much weaker than the LA branch. Between the $\Gamma$- and K-points, the TA mode is indistinguishable from the LA mode in Fig.~\ref{fig:cmp_TDS_phonon_disp}. The results of Plotkin-Swing et al. suggest, however, that the TA mode should indeed be visible given sufficient energy resolution \cite{plotkin-swing_hybrid_2020}. Comparing with Fig.~\ref{fig:cmp_spec_kpoints_collang3e+00mrad_thickness750}, the TA mode can be distinguished as a shoulder of the LA peak in our simulations. This means that the TA mode is not absent in the FRFPMS results, but it is comparatively weak and it is in consequence hard to discriminate it in Fig.~\ref{fig:cmp_TDS_phonon_disp}.

The ZA and ZO branches do not have a corresponding EELS intensity anywhere, which is expected, since the single scattering total momentum transfer $\mathbf{q}$ is essentially orthogonal to the direction of the incoming beam everywhere except at $\Gamma$-points in the diffraction plane and vibrational scattering is therefore expected to happen predominantly due to atomic displacements in the $x/y$-plane.

The visibility of the TO and LO modes in the EELS signal is highly dependent on the direction in the diffraction plane. Along the path $\Gamma$-M-$\Gamma$, only the LO branch is visible and the TO branch is not represented in the vibrational EELS. Along $\Gamma$-K-M-K-$\Gamma$ in Fig.~\ref{fig:cmp_TDS_phonon_disp}, the EELS signal does not show any intensity corresponding to the LO branch, but the TO branch gives rise to a strong signal within the K-M-K segment. Such directional dependence is included in the Born approximation treatment through a dependence on the scalar product $\mathbf{q}.\mathbf{e}$ of the momentum transfer $\mathbf{q}$ and phonon polarization vector $\mathbf{e}$ as already discussed in section~\ref{sec:results_spectral_change_as_function_of_detector_position_thickness}.

We observe furthermore weak vibrational EELS intensity corresponding to optical branches in the center of the diffraction plane, where the momentum transfer is approximately zero, i.e., $q\approx 0$, in both considered directions. In the same region, the background EELS intensity, i.e., the non-zero EELS signal observed in band-free regions of the dispersion relation, is much smaller than for other momentum transfers. Comparing this to the theoretical results of Senga et al., we note, that they predict a strong vibrational signal of the LO mode in this region reaching the central $\Gamma$-point \cite{senga_position_2019}. This difference can likely be attributed to dipolar scattering, which is not included in our current simulations. This interpretation is supported by their theoretical calculations for graphite, a non-polar material with a similar dispersion and structure as hBN, which also show the EELS signal of the LO mode not quite reaching the central $\Gamma$-point.

Lastly, we observe that the magnitude of the EELS of the LA branch exhibits an interesting behavior in the vicinity of different $\Gamma$-points. Near zero momentum transfer, the EELS signal is comparably low. At a momentum transfer of $\pm2.9$~\AA$^{-1}$ the EELS signal of the LA mode is much stronger, but weakens slightly around the $\Gamma$-point at $\pm5.8$~{\AA}$^{-1}$ before increasing further around the $\Gamma$-point at $\pm8.7$~{\AA}$^{-1}$. This behavior could hint at 
a dynamical diffraction effect, which distributes the signal in a non-trivial fashion around $\Gamma$-points.

Overall, given the precision of the MD force field, we observe a good qualitative agreement of the vibrational EELS computed by the FRFPMS method with the phonon dispersion as well as with the results of Plotkin-Swing et al.\ and Senga et al.\ at all points in the diffraction plane except for the $\Gamma$-points \cite{plotkin-swing_hybrid_2020,senga_position_2019}.

\section{Conclusions and Outlook}

In this article, we have reviewed the FRFPMS method for calculation of vibrational (STEM-)EELS. We have considered the effects of finite simulation box sizes and how these could potentially influence spectral shapes in conjunction with the $\delta$-thermostat. We have shown, how we can remedy this problem using a wider hotspot thermostat. This leads to reliable spectral shapes, yet comes at the cost of reduced energy resolution in the simulations. We have applied the developed machinery to vibrational EELS simulations of hBN assuming a parallel electron beam. From the generated data set, we have extracted a range of different results in order to illustrate how the FRFPMS method gives access to experimentally measurable signals.

The raw EELS calculated for large off-axis collection angle is a steeply decreasing function of energy, which exhibits peaks consistent with the peaks in the VPS of the system. Most of the steep background can be removed by multiplying the raw FRFPMS EELS by energy. The resulting spectrum shows fair agreement with the $x/y$-projected VPS. %

Spectral features as a function of the orientation of a detector with small collection angle exhibit features qualitatively consistent with expectations based on the phonon dispersion and simulation methods based on the first-order Born approximation. However, we have also seen that the FRFPMS EELS is collection angle dependent and thickness effects are also present. In the last step, we have shown how profiles through the FRFPMS diffraction pattern reveal the phonon dispersion relation in a way consistent with published experiments.

There are several open avenues for future research. One could focus on deeper understanding of the exact approximations at which the FRFPMS method operates. On the computational side the method itself could be improved upon by finding optimal settings for the hotspot thermostat or even by developing new thermostats, which would have improved properties over the hotspot thermostat, such as reduced tails. This is interconnected with the question of the nature of the background observed in the FRFPMS EELS and also the nature of the EELS at the $\Gamma$-points in the diffraction plane. Furthermore a more systematic study of thickness effects and dynamical diffraction would be of great interest, since we have observed here spectral changes as a function of thickness.

\section{Acknowledgements}
We acknowledge Swedish Research Council for financial support and Swedish National Infrastructure for computing (SNIC) at the NSC center (cluster Tetralith).

\bibliography{references,references_special}

\end{document}